\documentclass{PoS}

\title{The RHIC Beam Energy Scan Program: Results from the PHENIX Experiment}

\ShortTitle{PHENIX Beam Energy Scan Results}

\author{\speaker{J.T.~Mitchell (for the PHENIX Collaboration)}\\
        Brookhaven National Laboratory\\
        E-mail: \email{mitchell@bnl.gov}}

\abstract{
The PHENIX Experiment at RHIC has conducted a beam energy scan at several collision energies in order to search for signatures of the QCD critical point and the onset of deconfinement. PHENIX has conducted measurements of transverse energy production, muliplicity fluctuations, the skewness and kurtosis of net charge distributions, Hanbury-Brown Twiss correlations, charged hadron flow, and energy loss. The data analyzed to date show no significant indications of the presence of the critical point.
}

\FullConference{8th International Workshop on Critical Point and Onset of Deconfinement,\\
		March 11 to 15, 2013\\
		Napa, California, USA}

\begin{document}

\section{Introduction}
Recent lattice QCD calculations predict that there is a first order phase transition from hadronic matter to a Quark-Gluon Plasma that ends in a critical point. There is a continuous phase transition on the other side of the critical point. The Relativistic Heavy Ion Collider (RHIC) has conducted a program to probe different regions of the QCD phase diagram in the vicinity of the possible critical point with a beam energy scan. During 2010 and 2011, RHIC provided Au+Au collisions to PHENIX at $\sqrt{s_{NN}}$ = 200 GeV, 62.4 GeV, 39 GeV, 27 GeV, 19.6 GeV, and 7.7 GeV. The strategy of the data analysis focuses on looking for signs of the onset of deconfinement by comparing to results at the top RHIC energy, and searching for direct signatures of a critical point. Results from PHENIX covering charged particle multiplicity and transverse energy production, multipicity and net charge fluctuations, Hanbury-Brown Twiss correlations (HBT), charged hadron flow, and energy loss will be discussed.

\section{Multiplicity and Transverse Energy Production}
\begin{figure}[htbp]
\begin{center}
\includegraphics[width=0.5\textwidth]{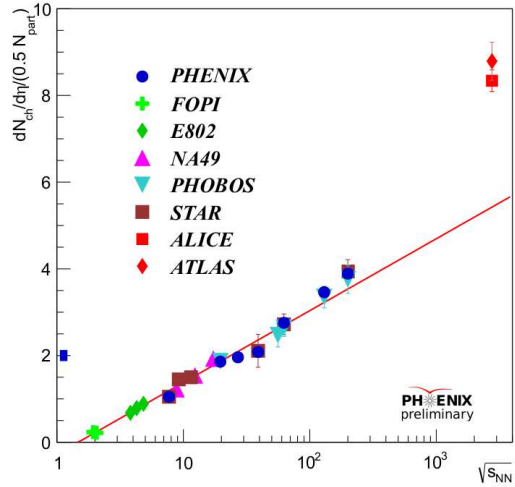}
\end{center}
\caption{The value of $dN_{ch}/d\eta$ at mid-rapidity normalized by the number of participant pairs as a function of $\sqrt{s_{NN}}$ for Au+Au collisions. The red line is an exponential fit to all of the data points excluding the ALICE and ATLAS points.}
\label{fig:nchExcite}
\end{figure}

PHENIX has measured charged particle multiplicity and transverse energy ($E_{T}$) production in Au+Au collisions at the following collision energies: 200, 62.4, 39, 19.6, and 7.7 GeV.  These observables are closely related to the geometry of the system and are fundamental measurements necessary to understand the global properties of the collision. This work extends the previous PHENIX measurements in 200, 130, and 19.6 GeV Au+Au collisions \cite{ppg019}. The charged particle multiplicity expressed as $dN_{ch}/d\eta$ normalized by the number of participant pairs is shown in Figure \ref{fig:nchExcite}. Included are measurements from other experiments including ALICE and ATLAS. The red line is a straight line fit to all of the points excluding the points at LHC energies. Charged particle production at LHC energies exceeds the trend established at lower energies.

\begin{figure}[htbp]
\begin{center}
\includegraphics[width=0.5\textwidth]{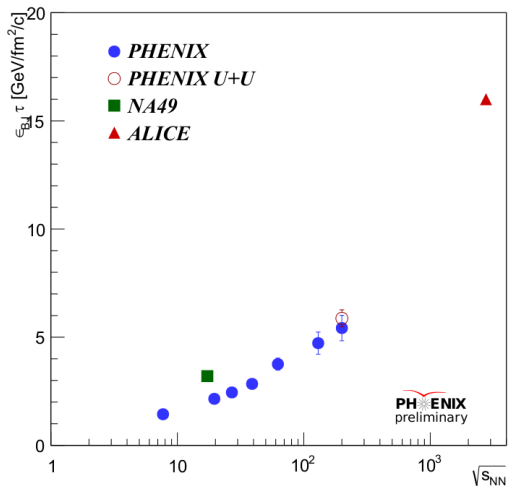}
\end{center}
\caption{The estimated value of the Bjorken energy density, $\epsilon_{BJ}$, multiplied by the formation time in central Au+Au collisions at mid-rapidity as a function of $\sqrt{s_{NN}}$. The open circle represents the estimate for 200 GeV U+U collisions.}
\label{fig:ebjExcite}
\end{figure}

Total $E_{T}$ production results are summarized in Figure \ref{fig:ebjExcite}, which shows the excitation function of the estimated value of the Bjorken energy density \cite{bjorken} expressed as
\begin{equation}
  \epsilon_{BJ} = \frac{1}{A_{\perp} \tau} \frac{dE_{T}}{dy},
\end{equation}
where $\tau$ is the formation time and $A_{\perp}$ is the transverse overlap area of the nuclei. The Bjorken energy density increases monotonically over the range of the RHIC beam energy scan. Also shown is the estimate for 200 GeV U+U collisions taken during the 2012 running period.

\begin{figure}[htbp]
\begin{center}
\includegraphics[width=0.5\textwidth]{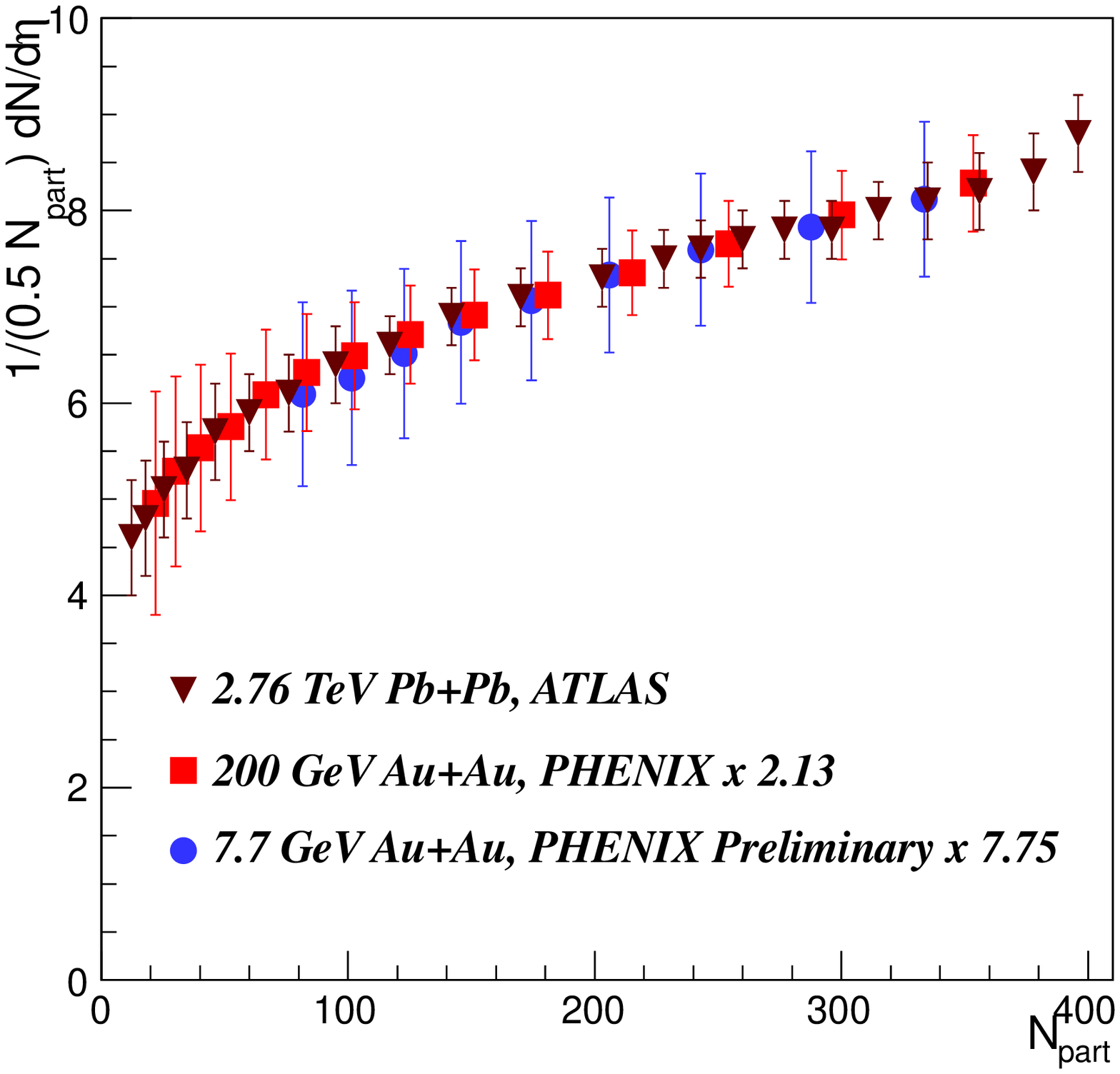}
\end{center}
\caption{$dN_{ch}/d\eta$ normalized by the number of participant pairs as a function $N_{part}$. Overlayed are the distributions from 7.7 GeV, 200 GeV, and 2.76 TeV Au+Au collisions. The PHENIX data has been scaled up to overlay the ATLAS data \cite{atlasNch}.}
\label{fig:dndetaRHICLHC}
\end{figure}

\begin{figure}[htbp]
\begin{center}
\includegraphics[width=0.5\textwidth]{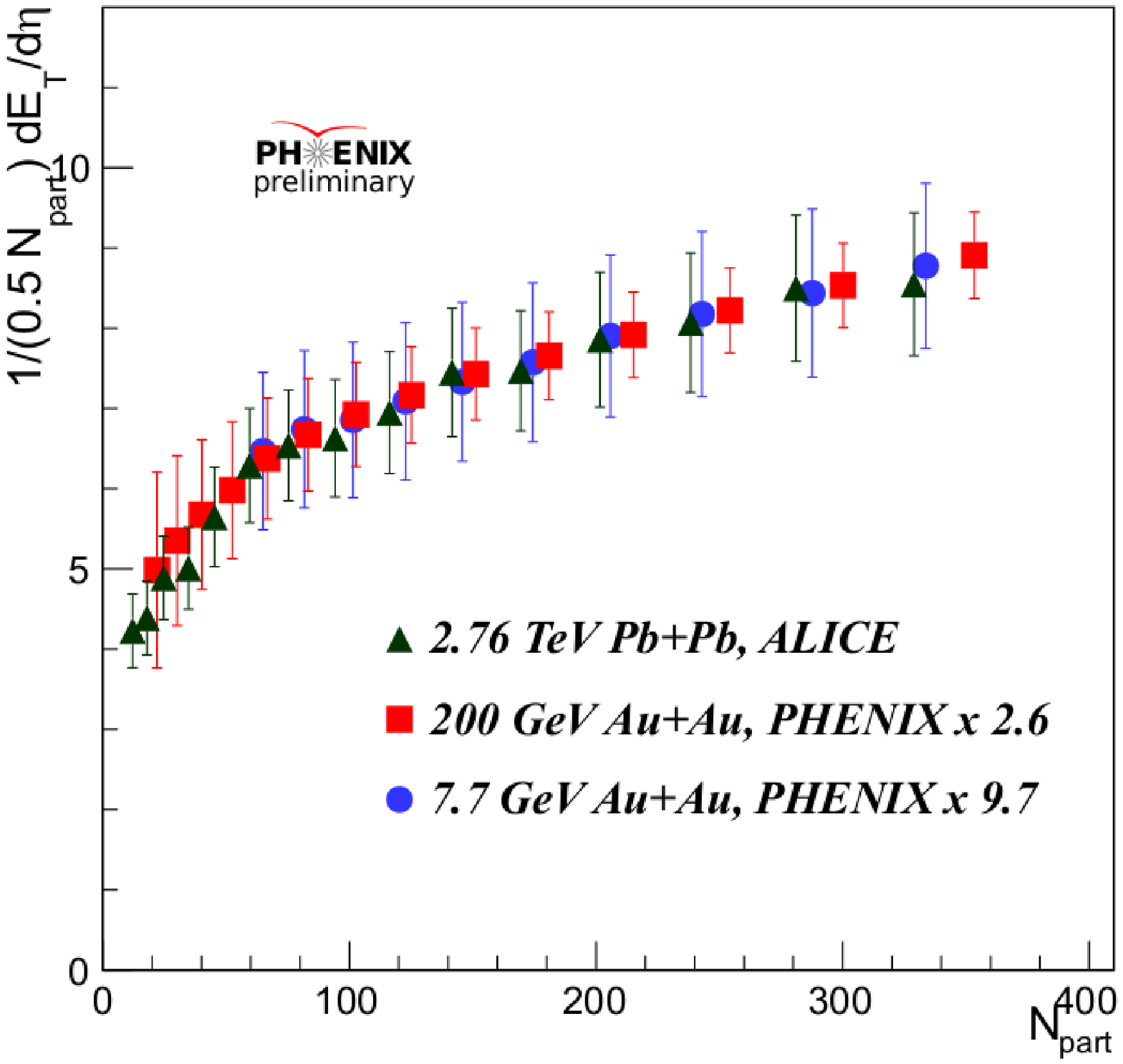}
\end{center}
\caption{$dE_{T}/d\eta$ normalized by the number of participant pairs as a function $N_{part}$. Overlayed are the distributions from 7.7 GeV, 200 GeV, and 2.76 TeV Au+Au collisions. The PHENIX data has been scaled up to overlay the ALICE data \cite{aliceEt}.}
\label{fig:detdetaRHICLHC}
\end{figure}

Although $N_{ch}$ and $E_{T}$ production dramatically increases at LHC energies compared to RHIC energies, the shape of the distributions as a function of the number of participants, $N_{part}$, is independent of the collision energy. This is illustrated in Figures \ref{fig:dndetaRHICLHC} and \ref{fig:detdetaRHICLHC}, which each show an overlay of the distributions for 7.7 GeV, 200 GeV, and 2.76 TeV Au+Au collisions. The 200 GeV and 7.7 GeV distributions have been scaled up to match the 2.76 TeV distributions. The shape of the distributions as a function of $N_{part}$ appears to be driven by the collision geometry.

\section{Multiplicity and Net Charge Fluctuations}

\begin{figure}[htbp]
\begin{center}
\includegraphics[width=0.5\textwidth]{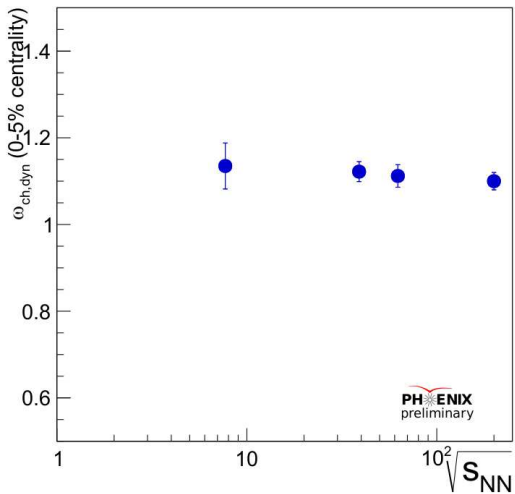}
\end{center}
\caption{Charged particle multiplicity fluctuations in central Au+Au collisions expressed in terms of the scaled variance as a function of $\sqrt{s_{NN}}$.}
\label{fig:svarExcite}
\end{figure}

Near the QCD critical point, it is expected that fluctuations in the charged particle multiplicity will increase \cite{Stephanov}. PHENIX has extended the previous analysis of multiplicity fluctuations in 200 and 62.4 GeV Au+Au collisions \cite{ppg070} to 39 and 7.7 GeV Au+Au collisions.  Charged particle multiplicity fluctuations are measured using the scaled variance, $\omega_{ch} = \sigma_{ch} / \mu_{ch}$, which is the standard deviation scaled by the mean of the distribution. The scaled variance is corrected for contributions due to non-dynamic impact parameter fluctuations using the method described in \cite{ppg070}. Figure \ref{fig:svarExcite} shows the PHENIX results for central collisions as a function of $\sqrt{s_{NN}}$. There is no indication of the presence of a critical point from the PHENIX results alone.

\begin{figure}[htbp]
\begin{center}
\includegraphics[width=0.5\textwidth]{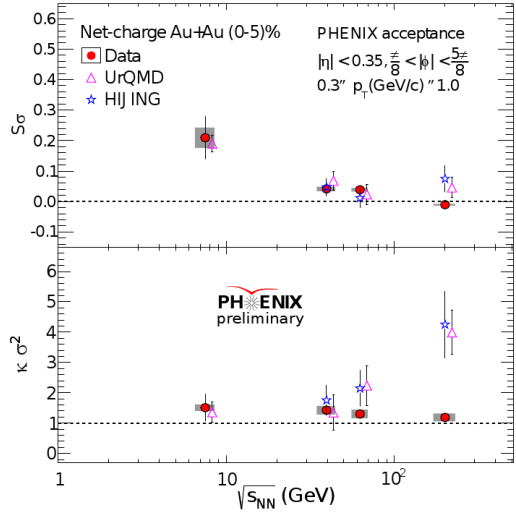}
\end{center}
\caption{The skewness multiplied by the standard deviation and the kurtosis multiplied by the variance from net charge distributions from central Au+Au collisions. The circles represent the data. The grey error bars represent the systematic errors. Also shown are URQMD and HIJING simulation results processed through the PHENIX acceptance. The increase in the kurtosis from URQMD and HIJING may be due to an increase in resonance production at 200 GeV.}
\label{fig:skewkurt}
\end{figure}

The shapes of the distributions of the event-by-event net charge are expected to be sensitive to the presence of the critical point \cite{Gavai}. PHENIX has measured the skewness ($S=\langle(N-\langle N \rangle)^{3}\rangle/\sigma^{3}$) and the kurtosis ($\kappa=\langle(N-\langle N \rangle)^{4}\rangle/\sigma^{4} - 3$) of net charge distributions in Au+Au collisions at 200, 62.4, 39, and 7.7 GeV.  These values are expressed in terms that can be associated with the quark number susceptibilities, $\chi$: $S\sigma \approx \chi^{(3)}/\chi^{(2)}$ and $\kappa\sigma^{2} \approx \chi^{(4)}/\chi^{(2)}$ \cite{Karsch}.  The skewness and kurtosis for central collisions are shown in Figure \ref{fig:skewkurt} as a function of $\sqrt{s_{NN}}$. The data are compared to URQMD and HIJING simulation results processed through the PHENIX acceptance and detector response. There is no excess above the simulation results observed in the data at these four collision energies. More details on this analysis are available in these proceedings \cite{phenixMoments}.

\section{Hanbury-Brown Twiss Correlations}

HBT measurements provide information about the space-time evolution of the particle emitting source in the collision. An emitting system which undergoes a strong first order phase transition is expected to demonstrate a much larger space-time extent than would be expected if the system had remained in the hadronic phase throughout the collision process \cite{Pratt1984}. The shape of the emission source function can also provide signals for a second order phase transition or proximity to the QCD critical point \cite{Csorgo}.

\begin{figure}[htbp]
\begin{center}
\includegraphics[width=0.5\textwidth]{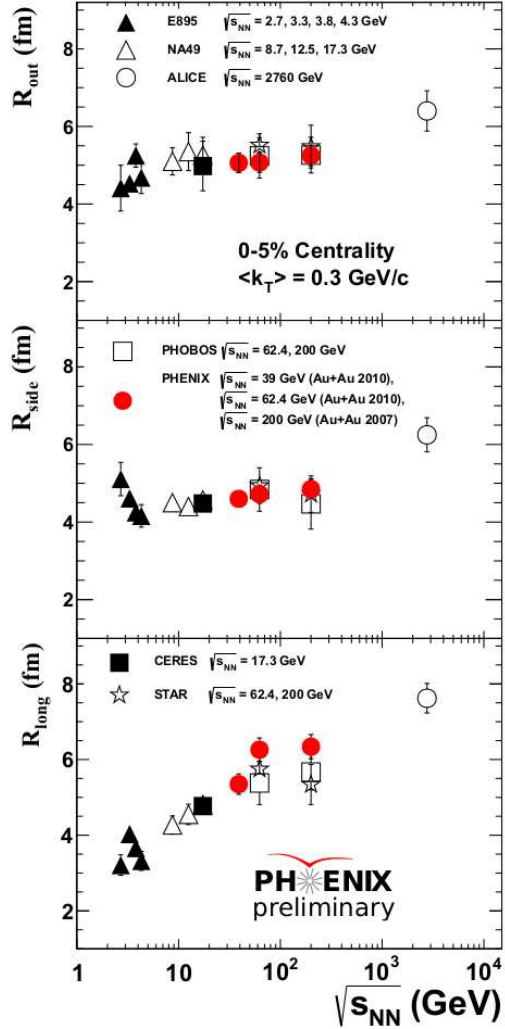}
\end{center}
\caption{HBT radii as a function of $\sqrt{s_{NN}}$ for central collisions at $<k_{T}>$=0.3 GeV/c. The red points are the PHENIX measurements. The data from other experiments can be found elsewhere \cite{e895HBT,ceresHBT,na49HBT,phobosHBT1,phobosHBT2,starHBT1,starHBT2,aliceHBT}.}
\label{fig:hbtRadii}
\end{figure}

\begin{figure}[htbp]
\begin{center}
\includegraphics[width=0.5\textwidth]{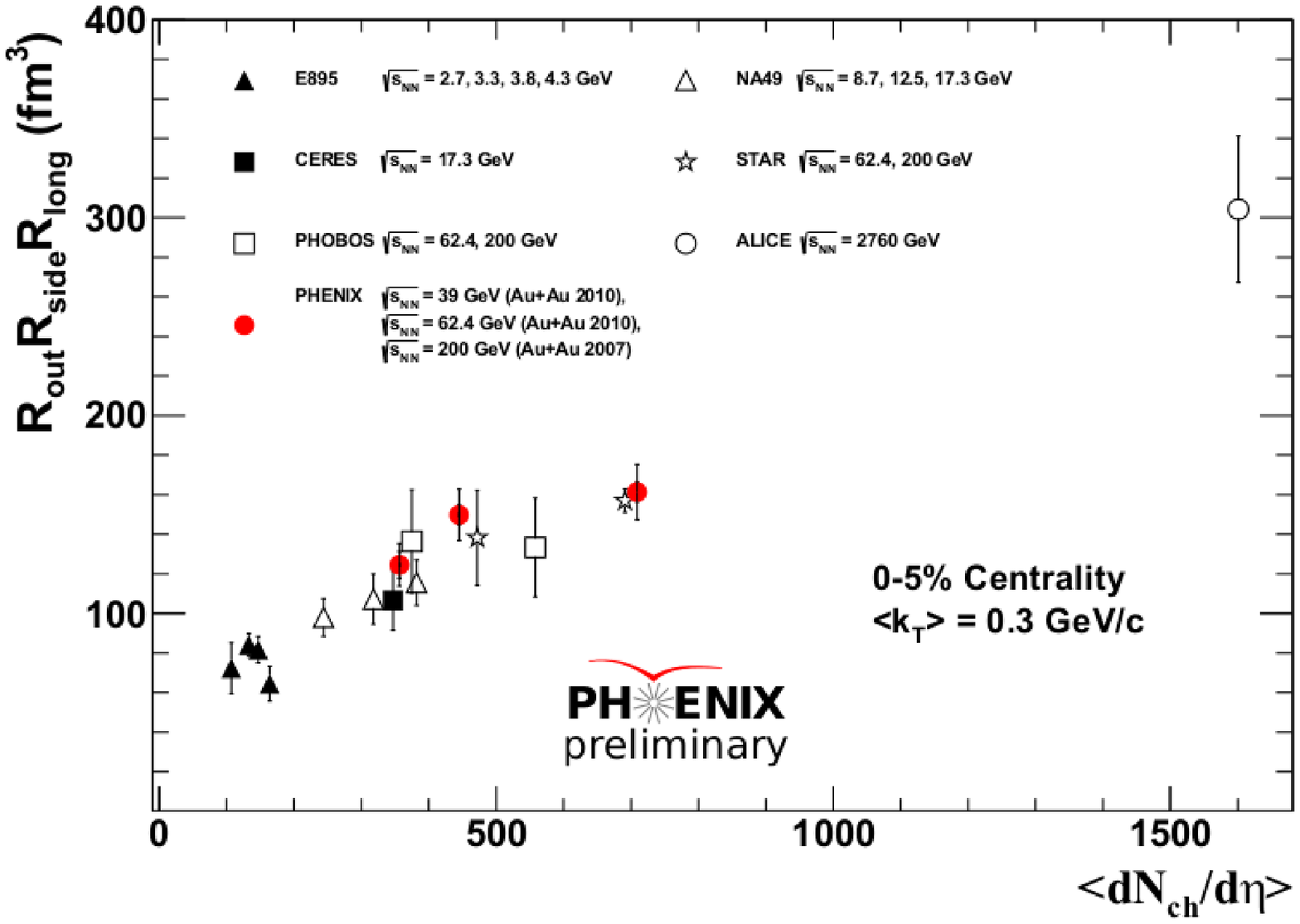}
\end{center}
\caption{The HBT freeze-out volume, $V_{f}$ as a function of $dN_{ch}/d\eta$ for central collisions at $<k_{T}>$=0.3 GeV/c. The red points are the PHENIX measurements. The data from other experiments can be found elsewhere \cite{e895HBT,ceresHBT,na49HBT,phobosHBT1,phobosHBT2,starHBT1,starHBT2,aliceHBT}.}
\label{fig:hbtVolume}
\end{figure}

PHENIX has measured the 3-dimensional source radii ($R_{side},R_{out}$, and $R_{long}$) for charged pions in 200 GeV, 62.4 GeV, and 39 GeV Au+Au collisions.  The measurements have been made for $0.2 < k_{T} < 2.0$ GeV/c. The results are summarized in Figure \ref{fig:hbtRadii}, which shows the excitation function for the radii for central collisions at $<k_{T}>$ = 0.3 GeV/c. There is no significant variation in the radii $R_{out}$ and $R_{side}$ over this energy range while $R_{long}$ follows an increasing trend as collision energy increases.  The freeze-out volume of the system can be estimated as follows: $V_{f} = R_{out} \times R_{side} \times R_{long}$. The excitation function of the freeze-out volume is shown in Figure \ref{fig:hbtVolume} as a function of $dN_{ch}/d\eta$.  From the lowest to the highest energies measured, the freeze-out volume increases linearly with the charged particle multiplicity.

\section{Charged Hadron Flow}

\begin{figure}[htbp]
\begin{center}
\includegraphics[width=0.5\textwidth]{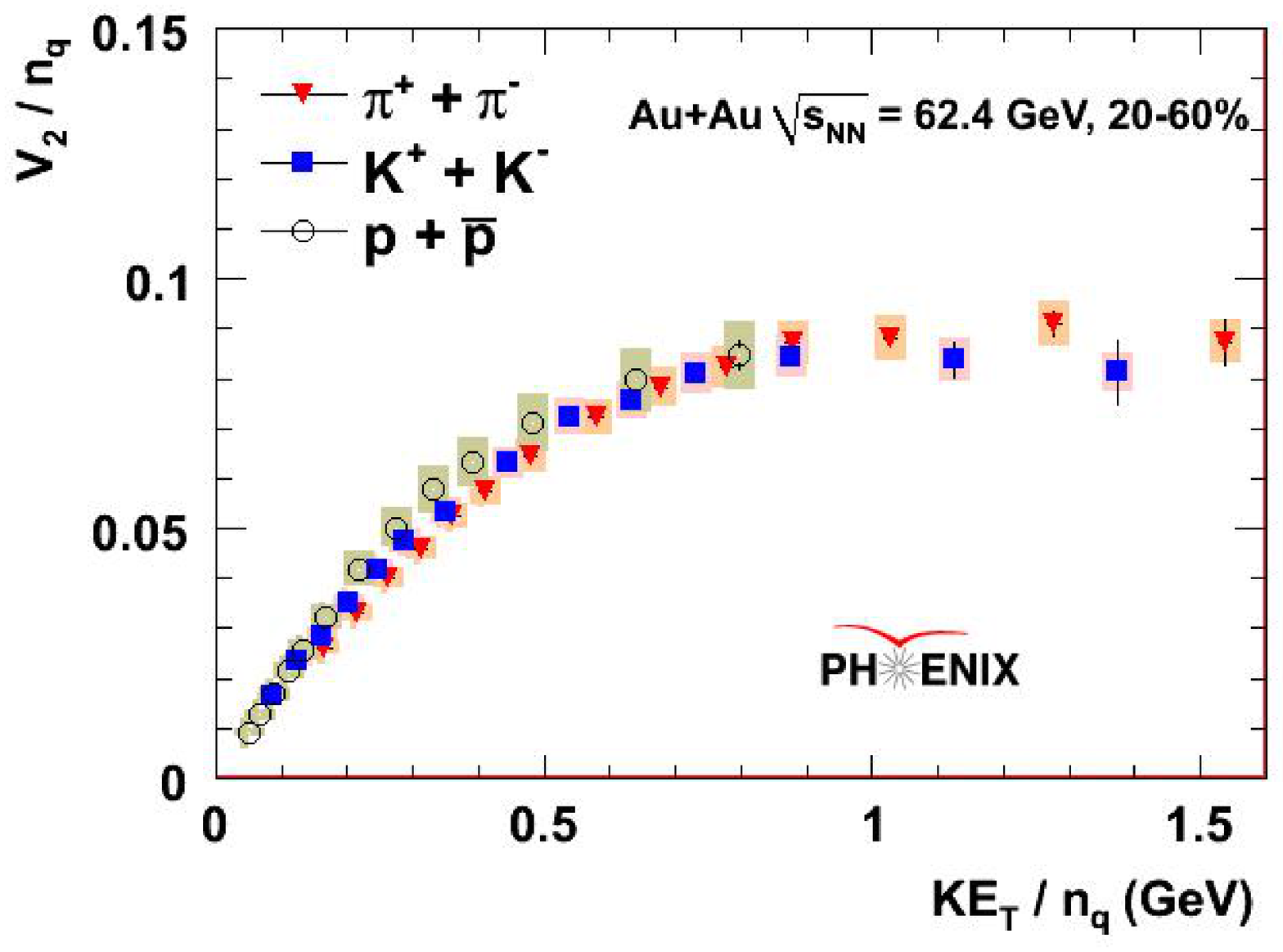}
\end{center}
\caption{The scaling of $v_{2}$ for 62.4 GeV Au+Au collisions. Shown are the measurements for identified pions, kaons, and protons.}
\label{fig:pidFlow62}
\end{figure}

\begin{figure}[htbp]
\begin{center}
\includegraphics[width=0.5\textwidth]{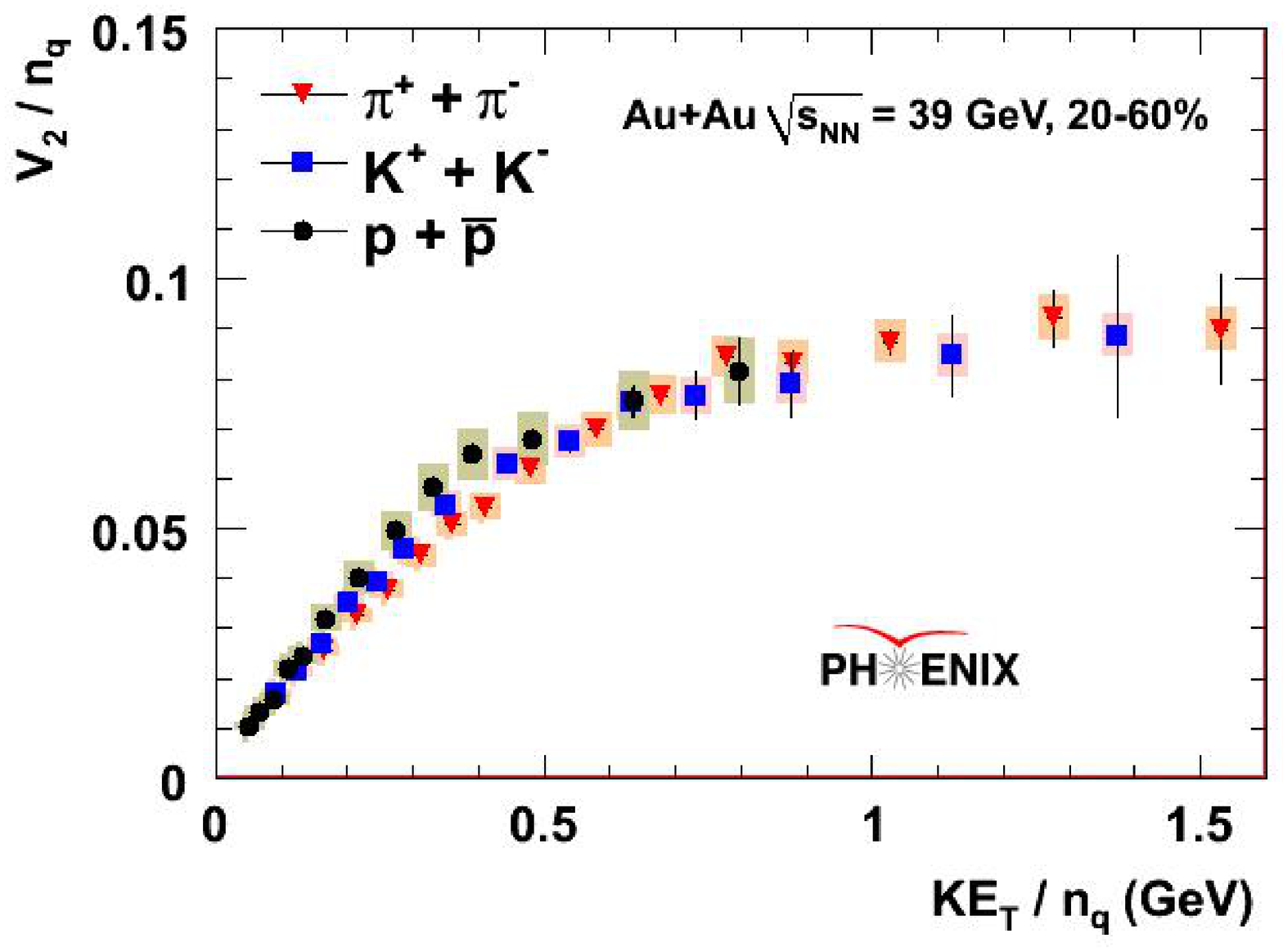}
\end{center}
\caption{The scaling of $v_{2}$ for 39 GeV Au+Au collisions. Shown are the measurements for identified pions, kaons, and protons.}
\label{fig:pidFlow39}
\end{figure}

Measurements of the anisotropy parameter $v_{2}$ for identified particles exhibit strong evidence of quark-like degrees of freedom at the top RHIC energies. A goal of the RHIC beam energy scan is to determine where the constituent quark scaling of $v_{2}$ no longer holds.  PHENIX has measured $v_{2}, v_{3}$, and $v_{4}$ for identified pions, kaons, and protons in 62.4 and 39 GeV Au+Au collisions. Shown in Figure \ref{fig:pidFlow62} and Figure \ref{fig:pidFlow39} are $v_{2}$ measurements scaled as $v_{2}/n_{q}^{n/2}$ on the vertical axis and $KE_{T}/n_{q}$ on the horizontal axis, where $n_{q}$ represents the number of quarks in the particle species being plotted, and $KE_{T}$ represents the transverse kinetic energy. At both of these collision energies, the scaling of $v_{2}$ observed at 200 GeV holds down to 39 GeV.

\section{Energy Loss}

\begin{figure}[htbp]
\begin{center}
\includegraphics[width=0.5\textwidth]{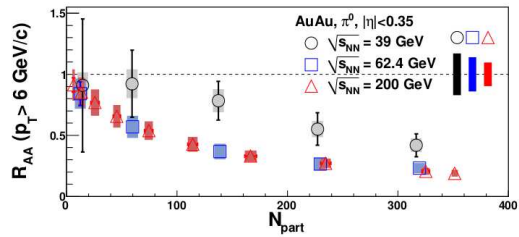}
\end{center}
\caption{The suppression factor, $R_{AA}$, for neutral pions with $p_{T} > 6$ GeV/c for 200, 62.4, and 39 GeV Au+Au collisions.}
\label{fig:pi0Raa}
\end{figure}

At the top RHIC energies, a very large suppression of hadron production at high transverse momentum is observed when compared to baseline p+p collisions \cite{phenixSuppress}. This suppression has been attributed to the dominance of parton energy loss in the medium. Previous studies of Cu+Cu collisions at $\sqrt{s_{NN}} = $ 200 GeV, 62.4 GeV, and 22.4 GeV \cite{phenixCuCu} show that suppression is observed (suppression factor $R_{AA} < 1$) at 200 and 62.4 GeV, but enhancement ($R_{AA} > 1$) dominates at all centralities at 22.4 GeV.  PHENIX has measured $R_{AA}$ for neutral pions in 200, 62.4, and 39 GeV Au+Au collisions \cite{phenixPi0Suppress}.  The value of $R_{AA}$ for neutral pions with $p_{T} > 6$ GeV/c is shown in Figure \ref{fig:pi0Raa} for all 3 energies.  There is still significant suppression observed in 39 GeV Au+Au collisions, but the suppression at the lower energy has decreased compared to the suppression seen in 62.4 and 200 GeV Au+Au collisions.  PHENIX has also measured the suppression of $J/\psi$ particles in 200, 62.4, and 39 GeV Au+Au collisions at forward rapidity \cite{phenixJpsiSuppress}.  Again, significant suppression is still observed in 39 GeV collisions, but the amount of suppression is decreased compared to that in 200 and 62.4 GeV collisions.

\begin{figure}[htbp]
\begin{center}
\includegraphics[width=0.5\textwidth]{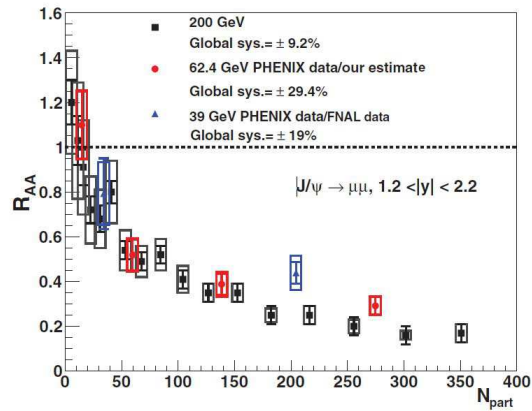}
\end{center}
\caption{The suppression factor, $R_{AA}$, for $J/\psi$ particles at forward rapidity for 200, 62.4, and 39 GeV Au+Au collisions.}
\label{fig:jpsiRaa}
\end{figure}

\section{Summary}
Presented here are some of the PHENIX results from the RHIC beam energy scan program. From the analyses completed to date, there is no significant indication of the presence of the QCD critical point.  Measurements of the suppression of neutral pions and $J/\psi$ particles suggest that the point at which the onset of deconfinement is seen may lie below collision energies of 39 GeV. Many analyses from PHENIX, particularly at $\sqrt{s_{NN}} = $ 27 GeV and 19.6 GeV, will be available soon.


\begin{thebibliography}{99}
\bibitem{ppg019} S.S.~Adler et al., Phys. Rev. C 71, 034908 (2005).
\bibitem{bjorken} J.~D.~Bjorken, Phys. Rev. D 27, 140 (1983).
\bibitem{atlasNch} G.~Aad et al., Phys. Lett. B 710, 363 (2012).
\bibitem{aliceEt} C.~Loizides et al., arXiv:1106.6324v1 (2011).
\bibitem{Stephanov} M.~Stephanov et al, Phys. Rev. D 60, 114028 (1999).
\bibitem{ppg070} A.~Adare et al, Phys. Rev. C 78, 044902 (2008).
\bibitem{Gavai} R.~V.~Gavai and S.~Gupta, Phys. Lett. B 696, 459 (2011).
\bibitem{Karsch} F.~Karsch and K.~Redlich, Phys. Lett. B 695, 136 (2011).
\bibitem{phenixMoments} P.~Garg et al., arXiv:1305.7327 (2013).
\bibitem{Pratt1984} S.~Pratt, Phys. Rev. Lett. 53, 1219 (1984).
\bibitem{Csorgo} T.~Csorgo et al., arXiv:nucl-th/0512060 (2005).
\bibitem{phenixSuppress} K.~Adcox et al., Phys. Rev. Lett. 88, 022301 (2001).
\bibitem{phenixCuCu} A.~Adare et al., Phys. Rev. Lett. 101, 162301 (2008).
\bibitem{e895HBT} M.~Lisa et al., Phys. Rev. Lett. 84, 2798 (2000).
\bibitem{ceresHBT} D.~Adamova et al., Nucl. Phys. A714, 124–144 (2003). 
\bibitem{na49HBT} C.~Alt et al., Phys. Rev. C77, 064908 (2008). 
\bibitem{phobosHBT1} B.~B.~Back et al., Phys. Rev. C73, 031901 (2006).
\bibitem{phobosHBT2} B.~B.~Back et al., Phys. Rev. C74, 021901, (2006).
\bibitem{starHBT1} B.~I.~Abelev et al., Phys. Rev. C79, 034909, (2009).
\bibitem{starHBT2} B.~I.~Abelev et al., Phys. Rev. C80, 024905, (2009).
\bibitem{aliceHBT} K.~Aamodt et al., Phys. Lett. B696: 328 (2011).
\bibitem{phenixPi0Suppress} A.~Adare et al., Phys. Rev. Lett. 109, 152301 (2012).
\bibitem{phenixJpsiSuppress} A.~Adare et al., Phys. Rev. C86, 064901 (2012).
\end{thebibliography}
\end{document}